\RequirePackage{fix-cm}
\documentclass[smallextended]{svjour3}       % onecolumn (second format)
\smartqed  % flush right qed marks, e.g. at end of proof
\usepackage{graphicx}
\begin{document}

\title{Studying gravitational deflection of light by Kiselev black hole via homotopy perturbation method%\thanks{Grants or other notes
}

\titlerunning{Studying gravitational deflection of light by Kiselev black hole...}        % if too long for running head

\author{V. K. Shchigolev        \and
        D. N. Bezbatko %etc.
}


\institute{V. K. Shchigolev \at
              Department of Theoretical Physics, Ulyanovsk State University,\\ 42 L. Tolstoy Str.,
Ulyanovsk, 432000, Russian Federation \\
            %  Tel.: +123-45-678910\\
             % Fax: +123-45-678910\\
              \email{vkshch@yahoo.com}           %  \\
           \and
           D. N. Bezbatko \at
              \email{bezbatko.dmitry@gmail.com}
}

\date{Received: date / Accepted: date}

\maketitle

\begin{abstract}
In this paper, the homotopy-perturbation method (HPM) is applied to obtain approximate analytical solutions for the gravitational  deflection of light in General Relativity near  Schwarzschild black hole surrounded by quintessence (Kiselev black hole). In order to demonstrate that HPM is able to yield acceptable solutions for the null-geodesics  with easily computable terms, the HPM is tested for the simple examples of spherically symmetric spacetimes such as  Schwarzschild  and  Reissner-Nordstr\"{o}m black holes. After that, the null-geodesics of light passing the vicinity of Kiselev black hole are studied  via the HPM in two particular cases regarding the equation of state parameter of quintessence. In addition,  a formula for the angle of deflection  has been obtained via HPM in the form of a series which  allows to calculate the angle with any accuracy without requirement of its smallness.
\end{abstract}

\section{Introduction}
\label{intro}
As well known, the gravitational deflection of light is one of the crucial predictions of the General Relativity (GR) and still plays a key role in understanding the problems related to Astronomy, Cosmology and Gravitational Physics \cite{Weinberg}.

Already Newton's theory of universal gravitation had predicted the deflection of path of light  due to the gravitational attraction. The Newtonian prediction for the deflection angle of light passing near a large mass $m$ is $\beta = 2Gm/(c^2l)$,
where $l$ is the closet distance of approach and approximately
the star radius. Later Einstein revealed that this  prediction was incorrect and the angular deflection should actually be twice greater of the predicted earlier. As known, this was subsequently confirmed by Eddington in 1919 through an experiment performed during the total solar eclipse (see, e.g., \cite{Will} and references therein).

In GR, the   deflection angle for a ray of light passing close to a gravitational mass can be obtained from the null geodesic, which the ray of light follows.
The exact analytical solution of light propagation even in Schwarzschild metric involves  elliptic integrals, but their evaluation is comparable with the efforts needed for the numerical integration of the geodesic equation \cite{Zschocke2}. Thus, approximate
analytical solutions of high accuracy are indispensable for the theoretical study as well as for the comparison with observational data. For example, the calculation of higher order deflection terms, due to
Schwarzschild black hole, from the null geodesic, has been performed recently in \cite{Iyer}.
The light deflection in Weyl conformal gravity was considered in \cite{Cattani} with the help of  the zeroth and first order linearized equations. At this, the integration of the linear equations was straightforwardly performed by using the standard perturbation method.

Basically, two approximative appraoches are used in order to determine the light deflection in weak gravitational fields.
The  first one is the standard parameterized post-Newtonian approach which applicable for $l >>m$, where $l$ is the impact parameter of the unperturbed light ray.
The second one is the standard weak-field approximative lens equation, which usually is
called the classical lens equations \cite{Bozza}. However, these exact lens equations are also given in terms of elliptic integrals. Therefore, approximations of these exact solutions are also needed for a time-efficient data
reduction. Several proposals for generalized
lens equations have then appeared in the literature. One decisive
advantage of the classical lens equation is its validity for arbitrarily small values of the impact distance $l$. A lens equation which allows an arbitrary large values of the deflection angle and used the deflection
angle expression for the Schwarzschild metric is obtained in \cite{Virbhadra}.

Up today, GR remains a very significant theory in modern physics and cosmology which is able to  give us new insights in our understanding
of gravity. Since the deflection of light as well as the perihelion precession are usually constrained in the solar system, it is
worthy to investigate both of them in the more general case (see, e.g. \cite{Hu}-\cite{Sumanta3} and references therein).

Moreover, the
observations of distant Ia-type supernova explosions indicate that starting at the cosmological redshift $z \approx 1$ expansion
of the Universe is accelerated \cite{Riess,Perlmutter}. Cosmologists proposed different models in order to explain this strange behavior of
the Universe such as the $\Lambda$CMD model (with a state parameter of $w = -1$) or dynamic scalar
fields. It is commonly accepted that this mysterious behavior comes from the existence of exotic dark energy.
Any modification of GR must be consistent with constraints astrophysical
scales as well as at the Solar system level. The same applies to the astrophysical models with a quintessence field.
For example,  the evolving quintessence scalar field dark energy
model and study the geodesics around a Schwarzschild black hole surrounded by
such scalar field was considered in \cite{Uniyal}.
The spherically symmetric solutions describing a black hole surrounded by dark energy in the form of a quintessential
field with equation of state in the form $p = w_q \rho$, with the quintessential parameter $-1 < w_q < -1/3$, has been found by Kiselev \cite{Kiselev}, and later has been investigated in several articles (e.g., in \cite{Jiao,Younas,Jamil}).

The idea of the Homotopy Perturbation Method which is a semi-analytical method was first proposed by  Ji-Huan He \cite{He,He2} for solving differential and integral equations. Later, the method is applied
to solve the non-linear and non-homogeneous partial differential equations.
The HPM has a significant advantage providing an analytical approximate solution to a wide range of
nonlinear problems of the fundamental and applied sciences \cite{Cveticanin}.
The HPM yields solutions in the form of rapidly
converging infinite series which can be effectively approximated by calculating only first
few terms. This method and a wide spectrum of its application have been substantially developed and studied for several years by numerous authors.
In contrast to the traditional perturbation methods, HPM does not require discretization, linearization or any restrictive assumption and small perturbations
in the equation to obtain an effective and simple solution.

Recently  there were studies in which this method was used for analytical computations in the field of cosmology and astrophysics (see, e.g. \cite{Shchigolev1} - \cite{Shchigolev3}.
It should be also mentioned that applications of HPM can be found in the field of astrophysics in different contexts that creates a new research field \cite{Aziz,Rahaman}.

In our paper, the  deflection of light is considered in
the 4-dimensional spherically symmetric spacetime by using HPM.
Two test problems are considered and the results  are compared with the two commonly acceptable results obtained earlier.
Our aim is not only to give one more application of HPM to the problem of light deflection in General Relativity, but also to obtain some new results for the light passing near Kiselev black hole. We discuss the null geodesics of the Kiselev black holes for two particular magnitudes of the quintessence EoS  when the
resulting formulae can be given in a relatively simple form.
Besides this, with the help of HPM,  we derive the simple formula for the angle of deflection  in the form of a series which  allows to calculate the angle with any accuracy without requirement of its smallness. However, it is worth noting the existence of other powerful methods to perform the same type of analysis. For example,  we can refer to one of such a method, the Adomian Decomposition Method, which was discussed in detail in \cite{Mak}.

\section{Formulation of the problem}

In this section, we give  the main equation of the null-geodesic motion in a spherical symmetry gravitational field in GR which are required to be solved for the deflection of light problem.
The stationary line element of 4-dimensional general spherically symmetric spacetime in GR \cite{Weinberg,Hu} can be represented by
\begin{equation}
ds^2=-f(r)dt^2+\frac{dr^2}{h(r)}+r^2(d\theta^2+\sin^2\theta d \varphi^2). \label{1}
\end{equation}

Since the path of light in GR is  treated as the null geodesic in spacetime, customarily one considers
the geodesics $\gamma(\tau)$ in the above spherically symmetric spacetime
expressed in the spherical coordinates $x^{\mu}=(t,r,\theta,\varphi)$ as $x^{\mu}(\tau)$, where $\tau$ is an affine parameter.

The geodesic $\gamma(\tau)$ can be obtained by solving the geodesic equation
$$
\frac{d^2x^{i}}{d\tau^2}+\Gamma^{i}_{kj}\frac{dx^{k}}{d\tau}\frac{dx^{j}}{d\tau}=0,
$$
where $\Gamma^{i}_{kj}$ are  the Christoffel symbols.

However, taking into account  the existence of Killing vectors, $\xi^a =(\partial/\partial t)^a$ and $\psi^a=(\partial/\partial \varphi)^a$, leading to such two conserved quantities as the total energy $\displaystyle E=f(r)\frac{dt}{d\tau}$ and the angular momentum $\displaystyle L=r^2\frac{d\varphi}{d\tau}$, and the constraint equation followed from (\ref{1}) along with $ds^2=0$, one can obtain the following equation for the coordinate $u\equiv1/r$
$$
\Big(\frac{d u}{d\varphi}\Big)^2=\frac{h(u)}{f(u)}\Big(\frac{E}{L}\Big)^2-h(u)u^2.
$$
Differentiating this equation with respect to $\varphi$, we get the second-order geodesic equation in metrics (\ref{1}) as follows
\begin{equation}
\frac{d^2 u}{d\varphi^2}=\frac{E^2}{2 L^2}\frac{d }{d u}\left[\frac{h(u)}{f(u)}\right]-h(u)u-\frac{1}{2}
u^2\frac{d h(u)}{d u}, \label{2}
\end{equation}
subject to the corresponding initial conditions for $u$ and $du/d\varphi$.

The main equation (\ref{2}) to be solved is a second-order nonlinear
differential equation. Our aim is to solve this equation analytically,
but in a certain approximation. Among all kinds
of approximate methods, we use here the HPM. The obvious advantage of this method is that there is no need to introduce a small parameter because it is contained in the method itself.

Since the HPM has now become standard, and for
brevity, the reader is referred to \cite{He,He2} for the basic
ideas of the HPM.  Considering  equation (\ref{2}) as the specific case of the following non-linear equation
$$
L(u) +  N(u)= 0
$$
for the function $u(\varphi)$, where $\varphi \in \Phi$, $L$ and $N$ are the linear and non-linear terms,  we construct a homotopy  $u(\varphi, p): \Phi\times [0,1] \to {\it I\!\!R} $  as follows
$$
H(u, p)=(1- p) [L(u)- L(u_0)]+p\,[L(u) +  N(u)] = 0,
$$
where   $p \in [0, 1]$ is an embedding parameter, and $u_0=u_0(\varphi)$ is an initial approximation. Hence, one can see that changing $p$ from $0$ to $1$ is the same as changing $H(u, p)$ from $L(u)- L(u_0)$ to $L(u) +  N(u)$, which are called homotopic.  By applying the perturbation procedure, we assume that the solution of (\ref{2})
can be expressed as a series in $p$, as follows:
\begin{equation}\label{3}
u(\varphi) = u_0(\varphi) + p\, u_1 (\varphi) + p^2 u_2(\varphi) + ...\, .
\end{equation}
When we put $p \to 1$, then equation $L(u) +  N(u)= 0$ corresponds to (\ref{2}), and (\ref{3}) becomes the approximate
solution of (\ref{2}), that is $u(\varphi)= \lim_{p \to 1} u =u_0(\varphi) + u_1 (\varphi) + u_2(\varphi)+ ...$.

\section{Test examples}

The following two examples demonstrate the use of HPM for the analytical computation of the deflection angle of light in the simplest spherically symmetric spacetimes (\ref{1}).

\subsection{Schwarzschild spacetime}

First, we shall consider the simplest case of metric (\ref{1}), namely,  the Schwarzschild spacetime describing the
gravitational field of an uncharged non-rotating star.
For the Schwarzschild solution, we have  $f(r)=h(r)=1-2m/r$, or
\begin{equation}
f(u)=h(u)=1-2mu. \label{4}
\end{equation}
where $m$ is the mass of a star.  Therefore,  equation (\ref{2}) for the null geodesic can be written as
\begin{equation}
\frac{d^2 u}{d\varphi^2}+u=3\,m\,u^2.
\label{5}
\end{equation}
In the absence of mass ($m=0$), the obvious analytic solution for (\ref{5}) is a straight
line expressed in polar coordinates
\begin{equation}
u(\varphi)= \frac{1}{l}\sin\varphi,
\label{6}
\end{equation}
where $l$ is a constant, that is the term $3m\, u^2$ comes from the correction of GR. Therefore, we can consider (\ref{6}) to be the null approximation for (\ref{5}).

Let us now consider the HPM for solving equation (\ref{5}). For this end, one can suppose the following homotopy
\begin{equation}\label{7}
u''+u-p\,\, 3 m u^2=0,\,\,\,\,\,\,p \in [0,1],
\end{equation}
where the prime denotes derivative with respect to $\varphi$. We assume that the solution of (\ref{5})
can be expressed as a series in $p$ by equation (\ref{3}).

According to (\ref{6}), the initial conditions for $u_0(0)$ and $u_i (0)$  can be chosen as follows
\begin{eqnarray}\label{8}
u_0(0)=0,~~~~u_0'(0)=\frac{1}{l},~~~~~~ \\
u_i(0)=u_i'(0)=0,~~~~~~~~~~~~~~\label{9}
\end{eqnarray}
where $i \geq 1$.
The substitution of (\ref{3}) into equation (\ref{7}) yields
\begin{eqnarray}
p^0&:&u_0''+u_0=0, ~~~~~~~~~~~~~~~~~~~~~~~~~~~~~~~~~  \label{10}\\
p^1&:&u_1''+u_1-3mu_0^2=0,\label{11}\\
p^2&:&u_2''+u_2-6mu_0 u_1=0, \label{12}\\
& &.\,\,.\,\,.\,\,.\,\,.\,\,.\,\,.\,\,.\,\,.\,\,.\,\,.\,\,. \nonumber
\end{eqnarray}
To solve the null geodesic equation (\ref{5}), one usually uses a perturbation
method based on the fact that the last term in this equation is much smaller
than the other terms due to the smallness of $m$. Therefore, to the first order in $p$ given by equation (\ref{11}), we get the same solution which follows from the standard perturbation method (see, e.g., \cite{Hu,Freire}). The essential difference of methods appears already in equation (\ref{12}).
\begin{figure}
\centerline{\includegraphics[width=8cm]{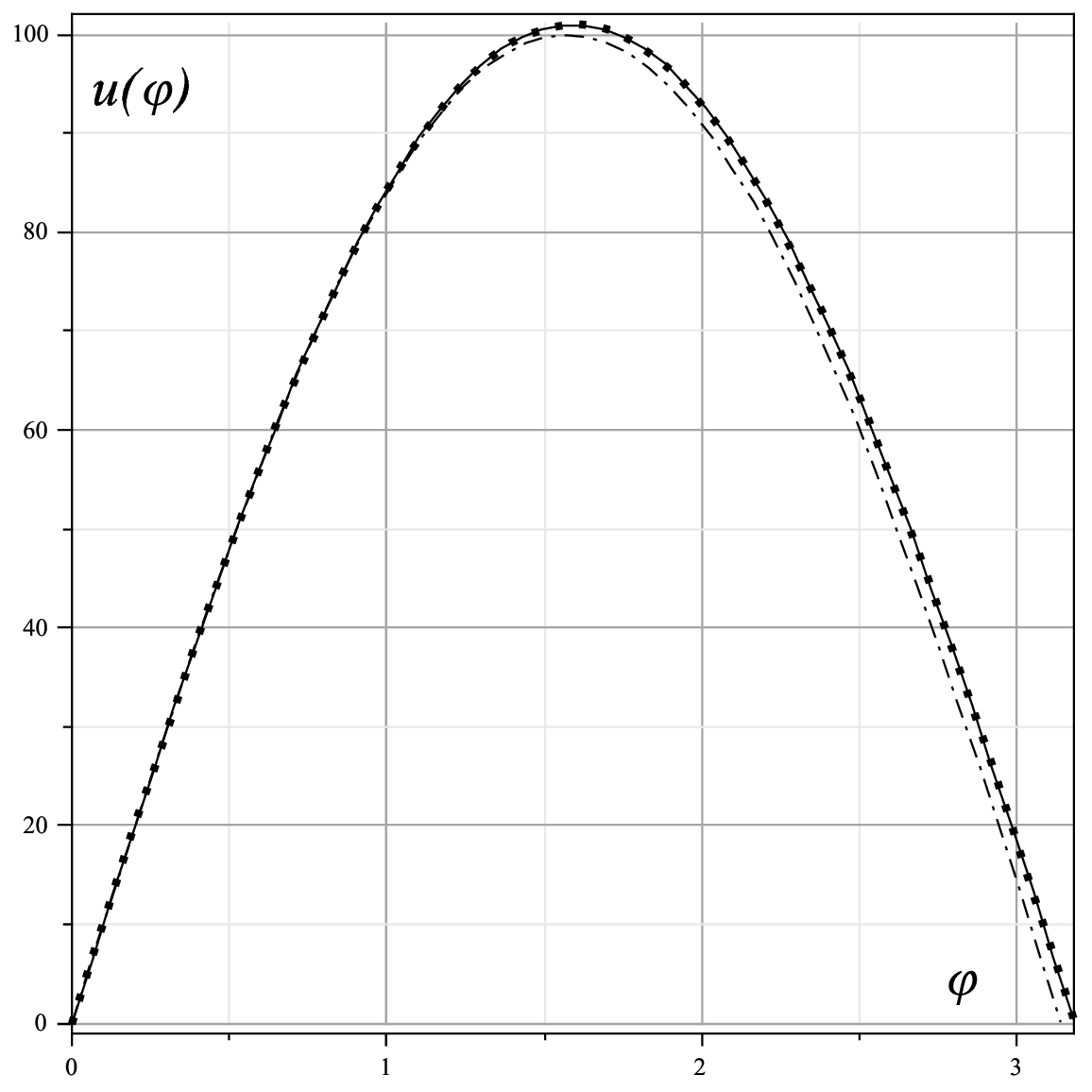}}
\vspace*{8pt}
\caption{Comparison of the approximate solutions, given by  Eq. (\ref{13})
(dot line), with the numerical solution to Eq. (\ref{5}) (solid line) and the  straight
line (unperturbed trajectory), given by Eq. (\ref{6}) (dash-dot line). Here, $m=0.0001$ and $l=0.01$ for the illustrative purpose. \label{f1}}
\end{figure}

It is noteworthy that we obtain the set of linear equations (\ref{10})-(\ref{12}). Their solutions with the initial conditions
(\ref{8}) and (\ref{9}) can be readily found. It yields the following approximate solution, $u\approx u_0+u_1+u_2$, for equation (\ref{9})
\begin{eqnarray}
u(\varphi)&=&\frac{1}{l}\sin\varphi+\frac{m}{l^2}\Big(1- \cos\varphi\Big)^2~~~~~~~~~~~~~~\nonumber\\
&+&\frac{m^2}{4l^3}\left\{2\sin\varphi-\left[\Big(3\cos\varphi-16\Big)\sin\varphi+
15\varphi\right]\cos\varphi\right\},\label{13}
\end{eqnarray}
where we have deliberately limited our calculation by the minimum degree of approximation.  All subsequent
approximations can also be obtained easily.

In order to demonstrate the accuracy of the method applied, the  graphs of $u(\varphi)$ with a certain choice of $m$ and $l$ for the numerical solution to  equation (\ref{5}) via the Maple package, and the approximate solution (\ref{13})  are given in Figure 1. Moreover, Table 1 shows the percentage of relative errors of the approximate solutions compared to the numerical one for the same example. There are serious reasons to expect that in other cases which we consider below,  the accuracy of the method may be of a similar order.

Obviously, solution (\ref{13}) satisfies  the initial condition $u(0) = 0$. Therefore, the deflection angle of light $\beta$ can be obtained from the equation
$u(\pi + \beta) = 0$,  using the small angle approximation
\begin{equation}
\sin (\pi + \beta) \approx -\beta,~~~~~ \cos (\pi + \beta) \approx -1.
\label{14}
\end{equation}
\begin{table}
\caption{The results of numerical solving with Maple and approximating (\ref{13})
(with the same $m$ and $l$ as in Fig. 1),  and the percentage of relative errors of the approximate solution.}
\label{tab:1}       % Give a unique label
\begin{tabular}{llll}
\hline\noalign{\smallskip}
$\varphi$ & Exact values $u(\varphi) from (\ref{5})$ & Approximate $u(\varphi) due to (\ref{13})$ & Errors \% of Eq.(13) \\
\noalign{\smallskip}\hline\noalign{\smallskip}
0.2 & 19.8673299727636774 & 19.86733042 & +2.2549$\times10^{-6}$\\
0.4 & 38.9480663023214576 & 38.94806617 & -3.3968$\times10^{-7}$\\
0.6 & 56.4947678749116876 & 56.49476426 & -6.3986$\times10^{-6}$\\
0.8 & 71.8276674296091784 & 71.82766143 & -8.3528$\times10^{-6}$\\
1.0 & 84.3587099036751909 & 84.35871006 & -1.8967$\times10^{-7}$\\
1.2 & 93.6114420917365920 & 93.61144174 & -3.7570$\times10^{-7}$\\
1.4 & 99.2364235428107833 & 99.23641776 & -5.8273$\times10^{-6}$\\
1.6 & 101.022212296170650 & 101.0221880 & -2.4050$\times10^{-5}$\\
1.8 & 98.9018403251726284 & 98.90176601 & -7.5140$\times10^{-5}$\\
2.0 & 92.9548323587580824 & 92.95465496 & -1.9084$\times10^{-4}$\\
2.2 & 83.4047925519158753 & 83.40443789 & -4.2523$\times10^{-4}$\\
2.4 & 70.6125828124728230 & 70.61191982 & -9.3892$\times10^{-4}$\\
2.6 & 55.0649260781590826 & 55.06378925 & -2.0645$\times10^{-3}$\\
2.8 & 37.3585896647602596 & 37.35679512 & -4.8036$\times10^{-3}$\\
3.0 & 18.1801476392888546 & 18.17752507 & -1.4443$\times10^{-2}$\\
\noalign{\smallskip}\hline
\end{tabular}
\end{table}
Thus, according to (\ref{13}), the angle is
\begin{equation}
\beta = \frac{4m}{l}\times\frac{\Big(1+\displaystyle\frac{15\pi }{16}\frac{m}{l}\Big)}{\Big(1-8\displaystyle\frac{m^2}{l^2}\Big)}\approx
\frac{4m}{l}\left(1+\frac{15\pi }{16}\frac{m}{l}+8\frac{m^2}{l^2}\right),
\label{15}
\end{equation}
which coincides with the usually used in GR angle $\beta=4m/l$ when the higher order terms $\mathcal{O}(m^2/l^2)$ are negligible. The same magnitude of $\beta$ could be obtained from (\ref{13}) while neglecting the last term in it. Nevertheless, the most interesting feature of our result  (\ref{15}) is the fact that the real deflection angle in Schwarzschild metric is slightly  greater than $\beta=4m/l$  \cite{Ellis}. As can be seen, the second-order correction $(15\pi/4)(m^2/l^2)$ to the angle of deflection that is discussed, for example, in article  \cite{Bodenner} here is obtained as a result of the minimal and simple calculation.

\subsection{Reissner-Nordstr\"{o}m spacetime}

In the case of the Reissner-Nordstr\"{o}m spacetime of a charged star, we have \cite{Weinberg}
\begin{equation}
f(u)=h(u)=1-2m u + Q^2 u^2, \label{16}
\end{equation}
where $Q$ is the charge. According to (\ref{16}), the null geodesic equation (\ref{2}) now becomes as follows
\begin{equation}
\frac{d^2 u}{d\varphi^2}+u=3m\,u^2-2Q^2u^3.
\label{17}
\end{equation}

Assuming that the unperturbed equation should have solution (\ref{6}), consider the following homotopy
\begin{equation}\label{18}
u''+u-p\,\big(3 m u^2-2Q^2u^3\big)=0,
\end{equation}
where  $p \in [0,1]$. Substituting (\ref{3}) into equation (\ref{18}), we obtain
\begin{eqnarray}
p^0&:&u_0''+u_0=0,~~~~~~~~~~~~~~~~~~~~~~~~~~~~~  \label{19}\\
p^1&:&u_1''+u_1-3mu_0^2+2Q^2u_0^3=0,\label{20}\\
p^2&:&u_2''+u_2-6mu_0u_1+6Q^2u_0^2u_1=0.\label{21}\\
& &.\,\,.\,\,.\,\,.\,\,.\,\,.\,\,.\,\,.\,\,.\,\,.\,\,.\,\,. \nonumber
\end{eqnarray}
The system of linear equations (\ref{19})-(\ref{21}) subject to initial conditions (\ref{8}), (\ref{9}) can be easily solved, giving
\begin{equation}
u(\varphi)=\frac{1}{l}\sin\varphi+\frac{m}{l^2}\Big(1- \cos\varphi\Big)^2
-\frac{Q^2}{4 l^3}\Big[ (\cos^2\varphi+2)\sin\varphi -3 \varphi \cos\varphi\Big]\label{22}
\end{equation}
for the simplest approximation of solution $u\approx u_0+u_1$.

Once again, the light deflection angle  $\beta$ can be obtained from the equation
$u(\pi + \beta) = 0$,  using the approximation (\ref{14}) in formula (\ref{22}). It leads to the following expression
\begin{equation}
\beta = \frac{4m}{l}-\frac{3\pi Q^2}{4l^2}.
\label{23}
\end{equation}
It should be noted that the deflection angle  (\ref{23}) was obtained from our solution (\ref{22}) only by the simple approximation (\ref{14}), while a similar expression in Ref. \cite{Hu} was obtained by a further simplification of the corresponding formula.

\section{Deflection of light  near Kiselev black hole}

Quintessence is the simplest scalar field dark energy model. The energy
density and the pressure of quintessence vary with time depending on the scalar field and the potential, which are
respectively given by:$ \rho_q = (1/2)\dot \phi^2+V (\phi)$ and $p_q = (1/2)\dot \phi ^2-V (\phi)$. One Schwarzschild-like solution related to the quintessence
model was found in \cite{Kiselev}.

The spacetime geometry of a static spherically symmetric black hole
surrounded by the quintessence (or Kiselev spacetime) is given by Eq. (\ref{6}) with \cite{Kiselev}
\begin{equation}
f(u)=h(u)=1-2m u -\sigma u^{3w_q+1}, \label{24}
\end{equation}
where $m$ is the mass of the black hole and $\sigma$ is the normalization factor satisfying
$0 < \sigma < 1$ \cite{Uniyal}. the quintessence equation of state parameter, $w_q=p_q/\rho_q$,  that is related to the energy density as follows $\rho_q=-3\sigma w_q u^{3(1+w_q)}/2$.

Substituting (\ref{24}) into equation (\ref{2}), one can obtain the following null geodesic equations in the gravitational field of Kiselev black hole
\begin{equation}\label{25}
\frac{d^2 u}{d\varphi^2}+u=3m\,u^2+\frac{3\sigma (w_q+1)}{2}u^{3w_q+2},
\end{equation}
which of course coincides with equation (\ref{5}) in the absence of quintessence, i.e. when $\sigma=0$. At $w_q \to -1$, the quintessence becomes  the cosmological constant $\Lambda=3\sigma/2$, and the last term in equation (\ref{25})  vanishes.

In this case, we construct the following homotopy for equation  (\ref{25}):
\begin{equation}\label{26}
\frac{d^2 u}{d\varphi^2}+u-p\left[3m\,u^2+\frac{3\sigma (w_q+1)}{2}u^{3w_q+2}\right]=0.
\end{equation}
Since  the numerical value of the exponent $(3w_q+2)$ in this equation is not specified yet, we have to express  $(u_0+pu_1+p^2 u_2+...)^{3w_q+2}$ in the form of Taylor series in $p$. Thus,
\begin{equation}\label{27}
u^{3w_q+2}=u_0^{3w_q+2}+p(3w_q+2)u_0^{3w_q+1}u_1+...
\end{equation}
Inserting (\ref{3}) and (\ref{27}) in equation (\ref{26}), we have the following system of linear equations
\begin{eqnarray}
p^0:~u_0''+u_0=0, ~~~~~~~~~~~~~~~~~~~~~~~~~~~~~~~~~~~~~~~~~~~~~~~~~~~~~~~~~\label{28}
\\
p^1:~u_1''+u_1-3mu_0^2-\frac{3}{2}\sigma (w_q+1) u_0^{3w_q+2}=0,~~~~~~~~~~~~~~~~~~\label{29}
\\
p^2:~u_2''+u_2-6mu_0u_1-\frac{3}{2}\sigma (w_q+1)(3w_q+2) u_0^{3w_q+1}u_1=0,\label{30}
\end{eqnarray}
subject to the initial conditions (\ref{8}) and (\ref{9}).

Unfortunately, equations (\ref{29}) and (\ref{30}) for an arbitrary value of the quintessence parameter can be integrated only in quadratures.
Therefore, we are going to
discuss in more detail the null geodesics of the Kiselev black holes with quintessential parameters $w_q = -1/3$ and $w_q = -2/3$ when the
resulting formulae can be given in a relatively simple form.

Let us first consider the case $w_q=-1/3$. With this EoS of quintessence, equations (\ref{28})-(\ref{30}) can be readily solved providing the following results
\begin{eqnarray}
u_0=\frac{1}{l}\sin\varphi,~~~~~~~~~~~~~~~~~~~~~~~~~~~~~~~~~~~~~~~~~~~~~~~~~~~\label{31}
\\
u_1=\frac{m}{l^2}(1-\cos\varphi)^2+\frac{\sigma}{2l}(\sin\varphi-\varphi\cos\varphi),~~~~~~~~~~~~~\label{32}
\\
u_2= \frac{1}{8l}\Big(3\sigma^2+4\frac{m^2}{l^2}\Big)\sin\varphi+\frac{2m\sigma}{l^2}(1-\cos\varphi)^2~~~~~~~~\nonumber \\
-\frac{1}{8l}\Big[6\frac{m^2}{l^2}\sin\varphi\cos^2\varphi+\Big(\sigma^2\varphi+\frac{8\sigma m}{l}\Big)\varphi\sin\varphi ~~~~\nonumber
\\
+3\Big(\sigma^2\!+\!10\frac{m^2}{l^2}\Big)\varphi\cos\varphi\!-
\!8\frac{m}{l}\Big(\sigma\varphi+4\frac{m}{l}\Big)\sin\varphi\cos\varphi\Big]\!.\label{33}
\end{eqnarray}

Using the simplest approximation, $u \approx u_0+u_1$, and equation (\ref{14}), one can obtain for the deflection angle
\begin{equation}\label{34}
\beta=\frac{4m}{l}+\frac{\pi \sigma}{2}.
\end{equation}
From this equation, it could be made a conclusion that an additional angle due to quintessence can reach up to $\pi/2$ at $\sigma \to 1$. Of course, it is not true since the approximate equation (\ref{14})  for the deflection angle $\beta$  is not valid in this case.
If we apply the same approximation (\ref{14}) for the approximate solution (\ref{31})-(\ref{33}) with  $u\approx u_0+u_1+u_2$, we obtain a cubic equation for $\beta$. Therefore, we assume that the values of $\beta^2$ and $\beta^3$ can be neglected compared to $\beta$. Then we get the following deflection angle
\begin{equation}\label{35}
\beta=\frac{\displaystyle\frac{4m}{l}+\frac{\pi \sigma}{2}+\frac{8m\sigma}{l}+\frac{3\pi}{8}\sigma^2+\frac{15\pi}{4}\frac{m^2}{l^2}}
{\displaystyle 1-\frac{\pi^2}{8}\sigma^2-2\pi\sigma\frac{m}{l}-8\frac{m^2}{l^2}}.
\end{equation}
The dependence of deflection angle (\ref{35}) regarding $m/l$ and $\sigma$ for some their specific values is graphically shown in Fig. 1. It is clear that the general character of this dependency compared with (\ref{34}) is kept.

\begin{figure}
\centerline{\includegraphics[width=8.7cm]{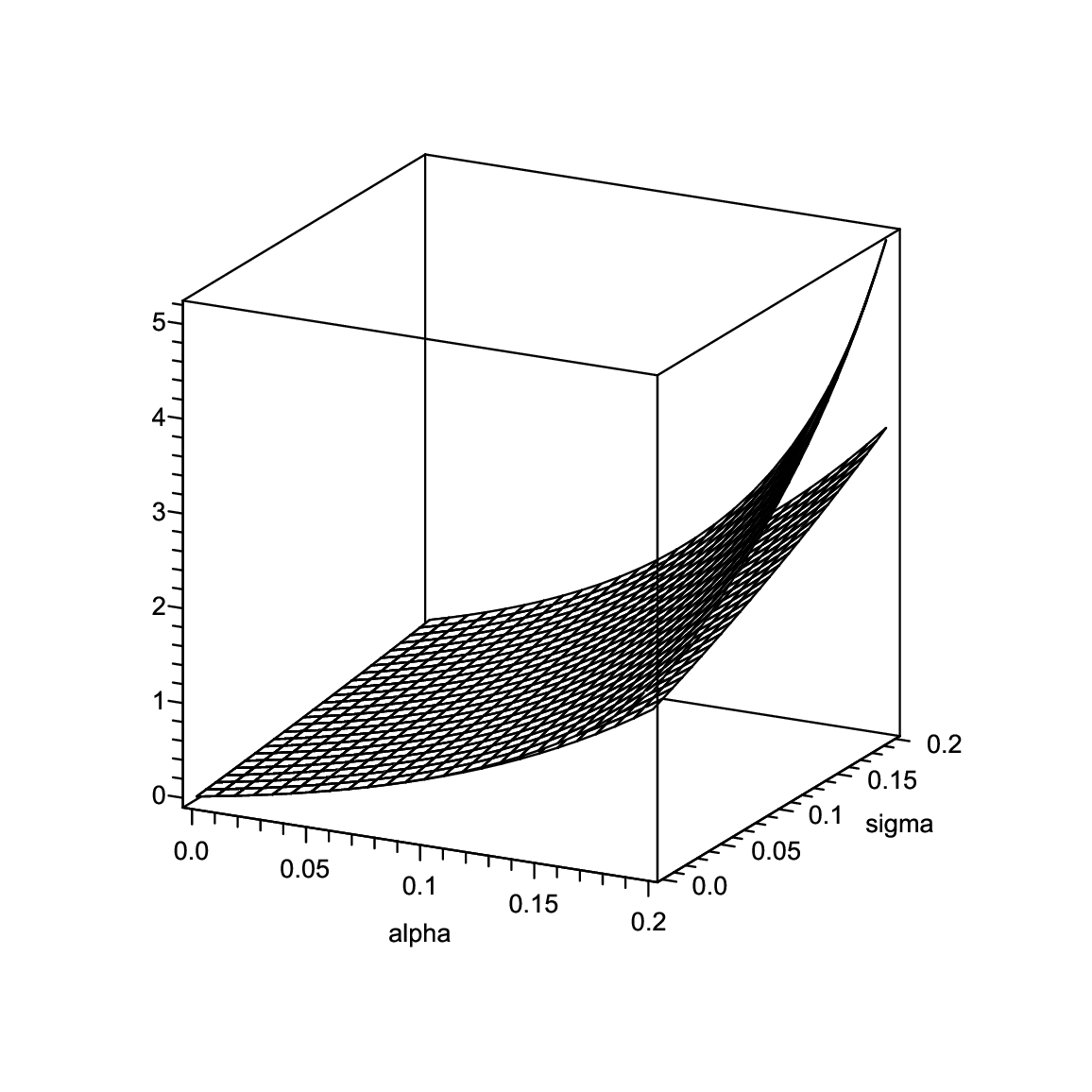}}
\vspace*{8pt}
\caption{The deflection angle $\beta$ versus $\alpha=m/l$ and $\sigma$ for equation (\ref{35}) (the upper graph) and equation (\ref{55}) (the lower graph).\label{f1}}
\end{figure}

In the case $w_q=-2/3$, equations (\ref{28})-(\ref{30}) become much simpler and can be solved easily, giving, for example,  $u\approx u_0+u_1+u_2$ by
\begin{eqnarray}
u(\varphi)=\frac{1}{l}\sin\varphi+\frac{m}{l^2}\Big(1- \cos\varphi\Big)^2+\frac{\sigma}{2}\Big(1- \cos\varphi\Big)~~~~~\nonumber\\
+\frac{m^2}{2 l^3}\sin\varphi+\frac{m\sigma}{2 l^2}\Big[(2\sin\varphi-3\varphi)\cos\varphi+\sin\varphi\Big]\nonumber \\
+\frac{m^2}{4l^3}\left[\Big(16-3\cos\varphi\Big)\sin\varphi-
15\varphi\right]\cos\varphi .\label{36}
\end{eqnarray}
Using the simplest approximation, $u\approx u_0+u_1$, and equation (\ref{14}), one can obtain the deflection angle
\begin{equation}\label{37}
\beta=\frac{4m}{l}+\sigma l.
\end{equation}
Comparing this angle with (\ref{34}), we can conclude that the additional deflection angle due to quintessence  nonlinearly varies  from $\pi\sigma/2$ at $w_q=-1/3$ to zero at $w_q=-1$. The more accurate approximation for $\beta$ can be obtained directly from equation (\ref{36}) as follows
\begin{equation}\label{38}
\beta=\!\!\left(\frac{4m}{l}+\sigma l+\frac{3\pi}{2}\frac{m}{l}\sigma+\frac{15\pi}{4}\frac{m^2}{l^2}\right)\!\!\!\left(1-2\sigma\frac{m}{l}\!-\!8\frac{m^2}{l^2}\right)^{-1}\!\!\!.
\end{equation}
One can easily note that both equations, (\ref{35}) and (\ref{38}) subject to $\sigma = 0$  give just the same value of the deflection angle $\beta$ as in the case of Schwarzschild black hole.
However, the most amazing feature of equations (\ref{37}) and (\ref{38}) is the appearance of the extra term $\sigma l$ which is proportional to the impact parameter $l$. This unusual  extra angle of deflection due to quintessence with EoS $w_q=-2/3$ requires a special discussion in a separate paper.

\section{Computation of deflection angle by HPM}

Since for finding the light deflection angle it is necessary to solve nonlinear equation $u(\varphi)=0$, in addition to the above used method,  again  it is useful to apply HPM. Indeed, let us consider the light-path equation as $u(\varphi)=(1/l)\sin\varphi+(1/l)U(\varphi)$, where the first term is the straight path of light without disturbing by gravity, that is $u_0(\varphi)$, and $U(\varphi)=l\,[u_1(\varphi)+u_2(\varphi)+...]$. Then, the deflection equation, $u(\varphi)=0$, becomes as follows
\begin{equation}\label{39}
\sin\varphi+U(\varphi)=0.
\end{equation}
Note that the correct approximate solution represented in the form (\ref{39}) has to satisfy $U(0)=0$.
Thus, we can construct the following homotopy
\begin{equation}\label{40}
\sin\varphi+q\, U(\varphi)=0,
\end{equation}
where $q \in [0,1]$ is a new embedding parameter. According to HPM, we assume that the solution of (\ref{39})
can be represented as a series in $q$, that is
\begin{equation}\label{41}
\varphi=\varphi_0+q\, \varphi_1+q^2 \varphi_2 + q^3 \varphi_3 + ...\, .
\end{equation}
At $q \to 1$, equation (\ref{40}) tends to (\ref{39}), and (\ref{41}) becomes the approximate
solution of (\ref{39}), that is $\varphi = \lim_{q \to 1} \varphi =\varphi_0+\varphi_1+\varphi_2 + \varphi_3 + ...$.
Next, we have to express  both terms in equation (\ref{40}) in the form of Taylor series in $q$ as
\begin{eqnarray}\label{42}
\sin\varphi\!=\!\sin\varphi_0\!+\!q\varphi_1\cos\varphi_0\!+\!q^2\!\!\left(
\varphi_2\cos\varphi_0\!-\!\frac{1}{2}\varphi_1^2\sin\varphi_0\!\right)\nonumber \\
+q^3\!\!\left(\varphi_3\cos\varphi_0\!-\!\varphi_1\varphi_2\sin\varphi_0\!-\!\frac{1}{6}\varphi_1^3\cos\varphi_0\!\right)\!\!+\!\!...
\end{eqnarray}
and
\begin{equation}\label{43}
U(\varphi)\!=\!U(\varphi_0)\!+\!q\varphi_1U'(\varphi_0) \!+\!q^2\!\!\left[\varphi_2U'(\varphi_0)\!+\!\frac{1}{2}\varphi_1^2U''(\varphi_0)\right]\!+\!...
\end{equation}
The substitution of (\ref{42}) and (\ref{43}) into equation (\ref{40}) yields the following set of simple algebraic equations
\begin{eqnarray}
q^0:~~\sin\varphi_0=0, ~~~~~~~~~~~~~~~~~~~~~~~~~~~~~~~~~~~~~~~~~~~~~~~~~~~~~~~~~~~~~~~~~~~~~~~~~\label{44}
\\
q^1:~~\varphi_1\cos\varphi_0+U(\varphi_0)=0,~~~~~~~~~~~~~~~~~~~~~~~~~~~~~~~~~~~~~~~~~~~~~~~~~~~~~~~~~~\label{45}
\\
q^2:~~\varphi_2\cos\varphi_0-\frac{1}{2}\varphi_1^2\sin\varphi_0+\varphi_1U'(\varphi_0)=0,~~~~~~~~~~~~~~~~~~~~~~~~~~~~~~~~~~~~~\label{46}
\\
q^3:~~\varphi_3\cos\varphi_0\!-\!\varphi_1\varphi_2\sin\varphi_0\!-\!\frac{1}{6}\varphi_1^3\cos\varphi_0
\!+\!\varphi_2U'(\varphi_0)\!+\!\frac{1}{2}\varphi_1^2U''(\varphi_0)=0,~~~~\label{47}
\\
.\,\,.\,\,.\,\,.\,\,.\,\,.\,\,.\,\,.\,\,.\,\,.\,\,.\,\,.\,\,.\,\,.\,\,.\,\,.\,\,.\,\,.\,\,.\,\,.\,\,.\,\,.\,\,.\,\,.\,\,.\,\,.\,\,.\,\,. ~~~~~~~~~~~~~~~~~~~~~~~\nonumber
\end{eqnarray}
Since we consider unperturbed path of light as the straight line, we take $\varphi_0=\pi$ for the solution to equation (\ref{44}). Therefore, the deflection angle is given by $\beta=\varphi_1+\varphi_2+\varphi_3+...$. Using equations (\ref{44})-(\ref{47}), one can obtain
\begin{equation}\label{48}
\beta\approx  U(\pi)\left(\!1+ U'(\pi)+\left[U'^2(\pi)\!+\!\frac{U(\pi)U''(\pi)}{2}\!+\!\frac{U^2(\pi)}{6}\right]\right).
\end{equation}
The obvious advantage of this formula is the absence of requirement for smallness of the angle. At the same time, the number of terms taken into account when calculating it according to this equation determines only the accuracy with which we find out this angle.
If necessary, the subsequent terms in the approximate formula (\ref{48}) can be easily obtained by the corresponding extension of series in equations (\ref{42}) and (\ref{43}). Note that the final accuracy with which the the angle $\beta$ can be find out from equation (\ref{48}) is determined also by the accuracy of the approximate equation (\ref{39}).

It should be emphasized that $U(\pi)$, $U'(\pi)$ and $U''(\pi)$ in (\ref{48}) cannot
be calculated with the help of equation (\ref{39}) since the latter is valid only for the specific values of angle, say, $\varphi =0$ and $\varphi = \pi + \beta$ subject to $\beta \neq 0$.

Moreover, we would like to note that one of the simplest methods  for finding successively better approximations to the roots of equation (\ref{39}), known as Newton's method, gives the following first-order approximation for the deflection angle
\begin{equation}\nonumber
\beta = \frac{U(\pi)}{1-U'(\pi)}.
\end{equation}
One can easily see that this equation gives the same value of angle as it represented by equation (\ref{48}) in the first-order approximation only if $|U'(\pi)|\ll 1$, that is when $(1-U'(\pi))^{-1}\approx 1+U'(\pi)$. Noteworthy that in the derivation of equation (\ref{48}), the restriction of such a kind is not needed.

Finally, if we have an exact solution for the null geodesic equation $u=u(\varphi)$ subject to $u(0)=0$ and the existence of
\begin{equation}\nonumber
\lim_{\varphi \to 0}\left[\frac{u(\varphi)}{\sin\varphi}\right]=l^{-1}<\infty,
\end{equation}
then we can also calculate the deflection angle according to (\ref{48}) introducing $U(\varphi)=l\,u(\varphi)-\sin\varphi$. At this, equation (\ref{48}) reduces to
\begin{equation}\label{49}
\beta\approx u(\pi)\Big(3\,l+ 3l^2\,u'(\pi) +l^3\Big[u'^2(\pi) + \frac{1}{2}u(\pi)u''(\pi)+\!\frac{1}{6}u^2(\pi)\Big]\Big).
\end{equation}
Taking the order of accuracy represented by equation (\ref{48}), let us specify the magnitudes of the deflection angles in the cases considered above.

\subsection{Deflection by Schwarzschild black hole}

Comparing our solution (\ref{13})  with (\ref{39}) in this case, one can get
\begin{equation}
U(\pi)=-U''(\pi)=4\frac{m}{l}+\frac{15\pi}{4}\,\frac{m^2}{l^2},\,\,U'(\pi)=8\frac{m^2}{l^2}.\label{50} \end{equation}
As one can see in (\ref{48}), the simplest approximation gives the well known formula, $\beta^{(0)}=U(\pi)=4m/l+(15\pi/4)(m^2/l^2)$. In the next approximation,
\begin{equation}
\beta^{(1)}=U(\pi)\Big(1+U'(\pi)\Big)=\left(4\frac{m}{l}+\frac{15\pi}{4}\frac{m^2}{l^2}\right)\left(1+8\frac{m^2}{l^2}\right),\nonumber \end{equation}
which almost coincides with (\ref{15}). Finally, the best approximation $\beta=\beta^{(3)}$ given by  (\ref{48}) is as follows
\begin{equation}
\beta=\left(4\frac{m}{l}+\frac{15\pi}{4}\frac{m^2}{l^2}\right)\left[1+\frac{8}{3}\frac{m^2}{l^2}-
10\pi\frac{m^3}{l^3}+\Big(64-\frac{75\pi^2}{16}\Big)\frac{m^4}{l^4}\right].\label{51} \end{equation}

\subsection{Deflection by Reissner-Nordstr\"{o}m black hole}

In this case, the comparison of (\ref{39}) with our solution (\ref{22}) yields
\begin{equation}
U(\pi)=-U''(\pi)=4\frac{m}{l}-\frac{3\pi}{4}\,\frac{Q^2}{l^2},\,\,\,\,\,\,\,\,\,U'(\pi)=0.\label{52} \end{equation}
Substituting (\ref{52}) into equation  (\ref{48}), one get the following angle of deflection in the gravitational field of Reissner-Nordstr\"{o}m black hole
\begin{equation}
\beta=\left(4\frac{m}{l}-\frac{3\pi}{4}\frac{Q^2}{l^2}\right)\left[1-\frac{16}{3}\frac{m^2}{l^2}+
2\pi\frac{mQ^2}{l^3}-\frac{3\pi^2}{16}\frac{Q^4}{l^4}\right].\label{53}
\end{equation}

\subsection{Deflection by Kiselev black hole}

In the case $w_q = -1/3$, by comparing the expressions (\ref{32}) and (\ref{33}) with
(\ref{39}) we obtain
\begin{eqnarray}\label{54}
U(\pi)=4\frac{m}{l}+\frac{\pi ~ \sigma}{2}+\frac{8m\sigma}{l}+\frac{3\pi}{8}\sigma^2+\frac{15\pi}{4}\frac{m^2}{l^2}, ~~~~~\nonumber \\
U'(\pi)=\frac{\pi^2}{8}\sigma^2+2\pi\sigma\frac{m}{l}+8\frac{m^2}{l^2},~~~~~~~~~~~~~~~~~~~~~~~~
\end{eqnarray}
and the corresponding deflection angle
(taking into account only the first derivative $U'(\pi)$) could be obtained  as follows
\begin{equation}\label{55}
\beta=\left(4\frac{m}{l}+\frac{\pi \sigma}{2}+\frac{8m\sigma}{l}+\frac{3\pi}{8}\sigma^2+\frac{15\pi}{4}\frac{m^2}{l^2}\right)
\left(1+\frac{\pi^2}{8}\sigma^2+2\pi\sigma\frac{m}{l}+8\frac{m^2}{l^2}\right).
\end{equation}
It is interesting to note that equation (\ref{35}) reproduces the same result for the deflection angle only if $U'(\pi)\ll 1$. A visual representation of the  difference between these two formulas is shown in Fig.1.

Comparing equation (\ref{36}) with (\ref{39}), we obtain
\begin{eqnarray}\label{56}
U(\pi)=\frac{4m}{l}+\sigma l+\frac{3\pi}{2}\frac{m}{l}\sigma+\frac{15\pi}{4}\frac{m^2}{l^2},~~~~~~~~~~~~ \nonumber
\\
U'(\pi)=2\sigma\frac{m}{l}+8\frac{m^2}{l^2}.~~~~~~~~~~~~~~~~~~~~~~~~~~~~~~~~~
\end{eqnarray}
Substituting (\ref{56}) into equation  (\ref{48}), one get the following angle of deflection in the gravitational field of Kiselev black hole with $w_q=-2/3$
\begin{equation}\label{57}
\beta = \left(\!\frac{4m}{l}+\sigma l+\frac{3\pi}{2}\frac{m}{l}\sigma+\frac{15\pi}{4}\frac{m^2}{l^2}\right)\left(
1+2\sigma\frac{m}{l}+8\frac{m^2}{l^2}\right)\!.
\end{equation}
It can be seen that equation (\ref{38}) yields the same result only if $2\sigma\frac{m}{l}+8\frac{m^2}{l^2}\ll 1$. Otherwise, the difference in these formulas will show itself similarly to that shown in Fig. 1 for the case $w_q=-1/3$.

\section{Conclusion}

In this article, HPM has been successfully applied to the solution of the differential equations for the light deflection in the gravitational field of black holes. Using this method, we have shown its advantages in obtaining the analytical approximations for the light deflection in the gravitational field. The results show the ease of application of the method and less computational complexity. With the help of simple calculations, in this article the higher-order approximations have been derived, including the light deflection in Kiselev space-time in two cases of EoS, viz. $w_=-1/3$ and $w_q=-2/3$. By applying HPM, we have shown that HPM is advantageous  in order to obtain the analytical approximate solutions of light deflection in GR. The obtained approximate solutions revealed that HPM is easy to implement in finding the analytical solutions and it reduces the size of the computational involvement compared to other popular methods.

Here, we deliberately gave up to discuss the choice of the homotopy, which may not be unique. We note only that accepted in this article homotopy greatly simplifies all calculations, but the solutions are represented by the series in the small physical parameters such as $m/l$ and $\sigma$. Taking into account a different choice of the homotopy, we can get rid of this lack, that we are going to demonstrate in our subsequent works.  Let us also note that the results are in a good agreement with the variational iteration method \cite{Shch,Shch1}.

In our view, Homotopy Perturbation Method offers excellent opportunity for the future
research. In this method, the analytic and approximate solutions are obtained without any restrictive assumptions for nonlinear terms as required by some existing techniques. Moreover, by solving some examples, one can see that the HPM appears to be very accurate to deal with reliable results.  The software used for the calculations in this study was Maple. The obtained results are of high accuracy even for the first-order approximate solution, showing that the solution procedure is acceptable. All this shows that HPM can be useful in various problems of astrophysics and cosmology.


\end{document}